\documentclass[journal=jacsat,manuscript=article]{achemso}

\input{settings.input}

\title{Energy Minimization of Discrete Protein Titration State Models Using Graph Theory}

\author{Emilie~Purvine}
\email{emilie.purvine@pnnl.gov}
\phone{+1 206 528 3461}
\affiliation{Computational and Statistical Analytics Division, Pacific Northwest National Laboratory}

\author{Kyle~Monson}
\email{kyle.monson@pnnl.gov}
\phone{+1 509 375 6379}
\affiliation{Computational and Statistical Analytics Division, Pacific Northwest National Laboratory}

\author{Elizabeth~Jurrus}
\email{elizabeth.jurrus@pnnl.gov}
\phone{+1 801 587 7880}
\fax{+1 509 375 2522}
\affiliation{Computational and Statistical Analytics Division, Pacific Northwest National Laboratory}

\author{Keith~Star}
\email{keith@pnnl.gov}
\phone{+1 509 372 4129}
\affiliation{Computational and Statistical Analytics Division, Pacific Northwest National Laboratory}

\author{Nathan~A.~Baker}
\email{nathan.baker@pnnl.gov}
\phone{+1 509 375 3997}
\affiliation{Advanced Computing, Mathematics, and Data Division, Pacific Northwest National Laboratory}
\alsoaffiliation{Division of Applied Mathematics, Brown University}

\begin{document}

\begin{abstract}
	There are several applications in computational biophysics which require the optimization of discrete interacting states; e.g., amino acid titration states, ligand oxidation states, or discrete rotamer angles.
	Such optimization can be very time-consuming as it scales exponentially in the number of sites to be optimized.
	In this paper, we describe a new polynomial-time algorithm for optimization of discrete states in macromolecular systems.
	This algorithm was adapted from image processing and uses techniques from discrete mathematics and graph theory to restate the optimization problem in terms of ``maximum flow-minimum cut'' graph analysis.
	The interaction energy graph, a graph in which vertices (amino acids) and edges (interactions) are weighted with their respective energies, is transformed into a flow network in which the value of the minimum cut in the network equals the minimum free energy of the protein, and the cut itself encodes the state that achieves the minimum free energy.
	Because of its deterministic nature and polynomial-time performance, this algorithm has the potential to allow for the ionization state of larger proteins to be discovered.
\end{abstract}

\maketitle

\section{Introduction}
There are many problems in computational physics and biophysics which require optimization over an exponentially large state space.
In this paper we demonstrate an algorithm adapted from computer vision for optimization over an exponentially large space in polynomial time for pairwise-decomposable interactions between states.
We focus on the problem of macromolecular titration state assignment; however, there are many other exponential space optimization problems in computational biophysics, including inverse folding \cite{godzik_novo_1993, yue_inverse_1992}, protein design \cite{dahiyat_novo_1997, gordon_energy_1999, richardson_novo_1989, samish_theoretical_2011}, and protein structure optimization \cite{ponder_tertiary_1987, ryu_protein_2013}.

Because hydrogens are rarely present in x-ray crystallographic structures, protein titration states often need to be computationally assigned to each titratable amino acid, including N- and C-termini, in the molecule \cite{antosiewicz_determinants_1996}.
{The bases for most modern protein \pkatext{} were established by Bashford and co-workers who developed both brute force and Monte Carlo procedures for generating titration curves \cite{BaDKaM1990, BaDKaM1991}}.
The \pkatext{} value of an amino acid or residue, together with the solution pH, provides a measure of the probability of protonation for a titration state: $\pka = -\log_{10} K_a$, where $K_a$ is the acid dissociation equilibrium constant for the residue.
Experimental methods provide the best mechanisms for determining both \pkatext{} values and titration states of a protein residue \cite{handloser_experimental_1973, mosher_proton_1972, reijenga_development_2013}, but experimental work is both time- and resource-consuming, so computational methods offer a compelling alternative for estimating \pkatext{} values and assigning titration states using a variety of physics- and statistics-based methods \cite{NeJGuMGaB2011}.
However, these calculations can be computationally demanding as they require the calculation of all $\order{N^2}$ pairwise interactions between all $N$ titratable residues, to determine intrinsic site \pkatext{} values \cite{yang_calculation_1993}, followed by optimization over the $\order{2^N}$ potential titration states of the system.

There are several approaches to such optimization, including sampling approaches such as Monte Carlo simulation {\cite{BaDKaM1990, BaDKaM1991, beroza_protonation_1991, LiZScH1987, metropolis_equation_2004, OzSMeH2004, ullmann_gmct_2012, wang_determining_2001}}, molecular dynamics \cite{AlBWaT1959, Kantardjiev2015, LiAKhMScH2005, MeJ2001}, {and} deterministic optimization methods such as dead-end elimination (DEE) \cite{DeJMaMHaBLaI1992, HaMKeDDoB2013}.
Sampling methods explore the optimization landscape using random move sets or trajectories generated from force-field-based equations of motion.
These methods have the advantage of generating thermodynamic ensembles through their sampling and are able to sample systems with complex energy functions; however, these methods are not guaranteed to find global minima.
In contrast, the DEE approach -- and its variants such as DEEPer \cite{HaMKeDDoB2013} -- are global optimization approaches. 
However, these approaches are restricted to pairwise-decomposable energy functions to accelerate the search, thus limiting the complexity of energy functions for the system.

The approach we describe in this paper is most closely related to DEE and its variants.
Like DEE, we are guaranteed to find a minimum energy state through a deterministic, polynomial time process.
The DEE algorithm scales cubically with the total number of rotamers in the system.
The algorithm we describe in this paper relies on the use of the max-flow/min-cut theorem.
There are new algorithms that approximate the minimum cut in roughly linear time in the number of edges \cite{chekuri_almost-linear-time_2014}, which results in an algorithm which is quadratic in the number of titratable residues.
As in DEE, we are currently limited by pairwise-decomposable energy functions, however with more research we hope to be able to extend this work to energy functions with higher order interactions (ternary, etc.), possibly through the use of hypergraphs.

\section{Methods} \label{sec:methods}
\subsection{Energy functions for titration state assignment}\label{sec:energy_funcs}
In this initial work, we will be following the simple and common approach \cite{yang_calculation_1993} of assigning titration states to a \emph{rigid protein}, wherein the backbone and amino acid locations are fixed.
{It is important to note that, for the current paper, we are assigning specific titration states and \emph{not} assigning \pkatext{} values \revision{or} titration probabilities.
Titration state assignment is important for setting up a variety of constant-titration calculations such as standard molecular dynamics simulations, docking simulations, etc.
The full calculation of titration curves is a longer-term application of this method.}
All titratable amino acids, with the exception of histidine (HIS), are assumed to have two possible states:  protonated or deprotonated.
This assumption ignores (or assumes equivalent) the various tautomers and conformers that can be present for many amino acids; these cases will be addressed in future work.
Histidine has three possible titration states which should not be considered equivalent \cite{Couch:2011aa}:  a singly protonated state with a hydrogen on N$_\epsilon$, a singly protonated state with a hydrogen on N$_\delta$, and a doubly protonated state with hydrogens on both nitrogens.
The state in which both N$_\delta$ and N$_\epsilon$ are deprotonated is highly energetically unfavorable and thus will be ignored.

For $N$ titratable residues, the set of all protonation states, $\pro$, of any protein without HIS can be described as the set of all $\{0, 1\}$ vectors of length $N$; i.e., $\pro = \{0,1\}^N$.
If there are $M$ HIS residues then $\pro = \{0,1\}^{N-M} \times \{0,1,2\}^M$.
Our goal in titration state assignment is to find \revision{a titration state $P$ in $\pro$} which minimizes the protein's energy
at a given pH (or proton activity), volume, and temperature $T$.
This free energy is often approximated as a pairwise-decomposable function between titration sites \cite{antosiewicz_protonation_2008, nielsen_chapter_2008}:
\begin{align} \label{energyFunc}
	\Etot(P) = \sum_{i=1}^N \gamma_i \ln(10) kT(\ph- \pkaint_i) + \sum_{i=1}^N \sum_{j=1}^N \gamma_i \gamma_j \Epair^{\intmath}(P_i, P_j)
\end{align}
where $\gamma_i$ is 1 when amino acid $i$ is charged and 0 otherwise, $\pkaint_i$ is the intrinsic \pkatext{} of amino acid $i$, and $\Epair^{int}(P_i, P_j)$ is the interaction energy between amino acids $i$ and $j$.
The intrinsic \pkatext{} value of amino acid $i$ is the \pkatext{} value if all other amino acids are in their neutral state.
Our formulation of the free energy is slightly different, though equivalent. 

Traditionally, the \pkatext{} of a given residue can be determined from the titration curve for that residue.
The titration curve is a plot of fractional proton occupancy vs. pH, and the pH value at which the fractional proton occupancy is 1/2 will give the \pkatext{} value.
When there is a greater than 0.5 probability that the residue is protonated (resp. deprotonated), then we consider it protonated (resp. deprotonated).
However, these fractional proton occupancies \cite{nielsen_chapter_2008} are computationally intensive to compute since they require computation of energy of the protein in all $2^N$ states:
\begin{align}\label{eqn:fi_frac_prot}
	f_i = \frac{\sum_{j=1}^{2^N} \gamma_i \exp{\frac{\Etot(P^j)}{kT}}}{\sum_{j=1}^{2^N} \exp{\frac{\Etot(P^j)}{kT}}},
\end{align}
where $f_i$ is the fractional proton occupancy of residue $i$, $j$ runs through all of the $2^N$ protonation states, $\gamma_i$ is 1 if residue $i$ is charged in state $j$ and zero otherwise, \revision{$P^j$ is the $j^{th}$ protonation state}, and $\Etot(P^j)$ is the energy described in \eqref{energyFunc} for protonation state $P^j$.

\revision{Instead of computing fractional occupancy via the ensemble average above, we follow basic two-state linkage analysis to approximate the protonated fraction \cite{wyman1990binding,popovic_modeling_2002}.
We define $G_{p}(i)$ to be the change in energy between the states where residue $i$ is protonated and where it is deprotonated.}
Given this definition, the fraction of protonated state is given as
\begin{align}
	\theta_p(i) &= \frac{e^{-\beta G_p(i)}}{ 1 \mathbf{+} a_H e^{-\beta G_p(i)}},
	\label{eqn:theta2}
\end{align}
where $\beta = \left( R T \right)^{-1}$, $R$ is the gas constant, $T$ is the temperature, and $a_H$ is the activity of the proton.
The form for HIS residues is similar but differs slightly due to the three-state system.
\revision{Below we drop the $i$ from the formulas when the residue is clear, for easier readability.}
If we assume the anionic state of HIS is energetically prohibited, then there are three potential states for the system
\begin{align}
	\theta_\delta  &= \frac{1}{1 + e^{-\beta \Delta G} + a_H e^{-\beta G_p}}, \\
	\theta_\epsilon &= \frac{e^{-\beta \Delta G}}{1 + e^{-\beta \Delta G} + a_H e^{-\beta G_p}}, \\
	\theta_p &= \frac{a_H e^{-\beta G_p}}{1 + e^{-\beta \Delta G} + a_H e^{-\beta G_p}},
	\label{eqn:theta3}
\end{align}
where $\Delta G$ is the difference in energy between the $\delta$ and $\epsilon$ states of HIS, $\theta_\delta$ is the fraction of states with HIS having a single proton on the $\delta$ nitrogen, $\theta_\epsilon$ is the fraction of states with HIS having a single proton on the $\epsilon$ nitrogen, and $\theta_p$ is the fraction of states with doubly protonated HIS.
Our method relies on finding a minimum energy state for the protein at any given pH value, changing the state of each residue, and recording the change in energy.
The method in this paper is focused on finding that minimum energy state efficiently.
We then use that information to calculate the titration curve using \eqref{eqn:theta2} and \eqref{eqn:theta3}, and subsequently the \pkatext{} value for each residue from the titration curves.

In order to calculate the individual energies contained in \eqref{energyFunc}, we employ PDB2PKA, a part of the PDB2PQR \cite{Dolinsky01072007, Dolinsky01072004} package based on the \pkatext{} methods of Nielsen et al.~\cite{nielsen_optimizing_2001}.
The interaction energy $\Epair(P_i, P_i)$ in \eqref{energyFunc} includes background and desolvation energies, while $\Epair(P_i, P_j)$ for $i \neq j$ includes Coulombic and steric interaction energies \cite{nielsen_optimizing_2001, Dolinsky01072007} between sites.
The background energy of site $i$ is simply the energy of the site if all other \revision{contributions} (other amino acids and the solvent) are removed. The desolvation energy, on the other hand, quantifies the interaction between the amino acid and the solvent, assuming all other amino acids are in their neutral state.
The electrostatic contributions to these energies are calculated via the Poisson-Boltzmann equation through the software package APBS \cite{Baker28082001}; the steric interaction energies are calculated via PDB2PQR \cite{Dolinsky01072007, Dolinsky01072004}. {When two amino acid states have steric clashes, a ``bump'' term is added in the form of a large unfavorable energy contribution to reduce the probability of the two states happening simultaneously}.
In the current work, the PARSE \cite{Tang20071475, SiDShKHoB1994} forcefield is used for protein radii and charge parameters.

{In addition to calculating interaction, background, and desolvation energies, PDB2PKA contains a Metropolis Monte Carlo for calculating titration curves and \pkatext{} values.
The PDB2PKA Monte Carlo algorithm starts with a random titration state for each residue in the protein.
For each step, the algorithm selects a random titration state for a random residue and calculates the energy difference $\Delta G_{i-1, i}$ with respect to the previous step.
If the energy is lower, the random titration move is accepted, otherwise it is accepted with a probability equal to $e^{\beta \Delta G_{i-1, i}}$.
The last 90\% of the steps in the Monte Carlo simulation are used to estimate \revision{fractional proton occupancy}
by calculating the fraction of Monte Carlo steps in which the residue was protonated.}

{In what follows, we compare this Monte Carlo approach with our energy minimization approach for the purpose of demonstrating our new optimization algorithms and qualitatively assessing their accuracy.
However, we recognize that these two algorithms use different approaches for \pkatext{} calculations.
The PDB2PKA Monte Carlo approach samples multiple states by implicitly calculating the probabilities presented in \eqref{eqn:fi_frac_prot}.
Our approach uses \revision{\eqref{eqn:theta2}} to directly calculate the \pkatext{} value, effectively neglecting the entropic contributions of multiple energetically accessible titration states.
In future work, we plan to improve our algorithm to use the calculated minimum energy state as the reference structure for Monte Carlo sampling.
}

\subsection{Discrete optimization and graph theory} \label{sec:compVis}
Research efforts in the field of computer vision (e.g., image restoration, image synthesis, image segmentation, multi-camera scene reconstruction, and medical imaging) focused on efficient algorithms for energy minimization via graph cuts in networks \cite{BoYVeOZaR2001,KoVZaR2004,VeO1999}.
Such applications often focused on the restoration of an  image (a collection of pixels and possible pixel labels); e.g., the image may contain noise to be removed, sections to be segmented, or disconnected images that need to be integrated into a single image.
These algorithms are designed to minimize an energy function which assigns an energy to each pixel based on its label (e.g., hue, intensity, segment membership, etc.) and to each pair of interacting pixels (usually neighbors) based on an interaction function.
Thus, the energy function takes the form
\begin{align} \label{eq:genForm}
	E(L) = \sum_{i=1}^N E_i(L_i) + \dsum_{(i,j) \in \mathcal{E}} E_{ij}(L_i,L_j)
\end{align}
where $L = \tup{L_1,L_2,\ldots,L_N}$, $L_i$ is the label of pixel $i$, and $\mathcal{E}$ is the list of pixel interactions, such that pixels $i$ and $j$ are said to interact if $(i,j) \in \mathcal{E}$.
Notice that our protein energy function \eqref{energyFunc} can be written in this form with
\begin{align*}
	E_i    &= \gamma_i \ln(10) kT (pH-\pkaint_i)+\gamma_i^2 \Epair^{\intmath}(P_i, P_i),\\
	E_{ij} &= \gamma_i \gamma_j \Epair^{\intmath}(P_i, P_j),\\
    \mathcal{E} &= \{ (i,j) : i \neq j\}.
\end{align*}
As a result of this similar pairwise-interaction form, we can apply the discrete minimization techniques established for computer vision to our protein problem.
In the case of \revision{two-state labels}, where amino acids can be only protonated or deprotonated, we can use these computer vision optimization methods directly.
We are also able to adapt these methods to the case of HIS which has three titration states, as described \revision{below}.

\subsection{Application of graph theory to energy minimization}  \label{sec:energyGraph}
To begin, we restrict our attention to proteins without HIS so that each amino acid has only two choices for its ``label'':  protonated or deprotonated.
\revision{Below}, we will discuss a method to include HIS in the \revision{graph-cut algorithm described above}.
The first step is to create a \revision{weighted} graph which holds all of the information from the energy function.
\revision{We will call this the \emph{energy graph}.}
Each amino acid will be represented by two vertices, one for each titration state.
If there are $N$ amino acids in the protein then there are $2N$ vertices in the graph.
Each $\langle$amino acid, configuration$\rangle$ pair, $\tup{i,P_i}$, has energy contributions from the difference between its intrinsic $\pka$ and the current pH value,
\revision{as well as the $\gamma_i^2 \Epair^{\intmath}(P_i, P_i)$ background and desolvation energies,}
which we combine into $E_i(P_i)$ as in \eqref{eq:genForm}.
This energy will be assigned as a weight on each vertex.
Additionally, each pair of amino acids can interact in both of their titration states, but amino acid $i$ in \revision{its deprotonated state} cannot interact with itself in \revision{its protonated state}.
Therefore, for each pair of amino acids there are 4 edges,
with a total of $4 {N\choose 2}=2N(N-1)$ edges for the entire protein.
Edge weights are given by the interaction energy for the particular amino acids and configurations, \revision{$\gamma_i \gamma_j \Epair^{\intmath}(P_i, P_j)$} as in \eqref{energyFunc}.
\revision{Additional information including graph theory background, details for constructing the energy graph, and example graphs are given in Supporting Information.}

The energy graph can be simplified by moving some edge weights to the vertices, and some vertex weights into a universal constant.
Through this simplifying process, we will end up with the \emph{normal form} of the energy graph \cite{KoVRoC2007} where all of the edge and vertex weights are non-negative and as many as possible have weight zero.
The normal form energy graph is \revision{then} transformed into an \emph{energy flow network} using a procedure which guarantees that the minimum cut in the network will equal the minimum energy of the protein titration system.
The graph transformation via simplification to normal form and representation as \revision{an energy} flow network is described in \revision{Supporting Information}.
Note that, after these transformation processes, the edge weights still represent energy values but are no longer interaction energies between the protonation states represented by the vertices in the edge.
\revision{Instead, these energies represent interactions between groups in the new sparser graph created by the normalization process.}


\revision{The energy of a protonation state can be determined by choosing the vertices in the  energy flow network corresponding to that particular protonation state, discarding all other vertices (and corresponding edges), and taking the sum of the edge and vertex weights that remain.
Therefore, a na\"{\i}ve approach to energy minimization would use this procedure to find the energy for all $2^N$ possible protonation states and choose the state which has the minimum energy.
However, this brute-force algorithm has exponential complexity, making it infeasible for even moderate-size proteins.
This selection of vertices and edges in the energy graph can also be represented through a graph cut in the energy flow network as detailed in Supporting Information.
Briefly, a graph cut defines a given protonation state of the protein by assigning all vertices associated with that protonation state into one set, and all others into a second set.
It can be shown that the cut value associated with this cut, the sum of edge weights for all edges from vertices in the first set to vertices in the second, plus the global constant from the normal form procedure, is exactly the energy of the associated state of the system \cite{KoVRoC2007}.
Additional requirements are needed to ensure that the minimum cut in the network yields the minimum energy configuration.
In a 2004 paper, Kolmogorov and Zabih \cite{KoVZaR2004} proved for pairwise-interacting states where all pairwise interactions are \emph{submodular}, it is possible to find the exact minimum energy state in polynomial time by computing the minimum cut on the flow network of the associated graph.
For our titration state application, submodularity means that, for all pairs of amino acids, the interaction energy when they are in the same state is smaller than the interaction energy if they are in different states.
In other words, submodularity implies that the sum of interaction energies for the singly-protonated states exceeds the sum of those for the the doubly-deprotonated and doubly-protonated states.}

\revision{Protein titration site interaction networks are not guaranteed to have submodular energies.
However, it is still possible to use the graph-cut method to label the portion of the amino acids whose energy functions are submodular \cite{KoVRoC2007}.
The remaining amino acids must be assigned by some other optimization method (e.g., Monte Carlo or brute force) on only the unassigned amino acids.
In the Discussion, we discuss the practical implications of unassigned amino acids.
\revision{We note that the number of unassigned residues is highly dependent on algorithms used to construct and cut the interaction energy graph.
Because of this issue,} we are not able to predict the number of unassigned residues by just looking at the interaction energies.
In future work, we plan to exploit the use of multiple different minimum cut algorithms to obtain the smallest number of unassigned amino acids, since all are fast to run but some yield fewer unassigned amino acids than others.}

\subsection{Moving beyond two states per amino acid}\label{sec:HISfix}
The discussion of the previous section was limited to minimizing \revision{two-state titratable systems} using graph cuts.
However, as mentioned in the introduction, not all titration sites can be represented simply by two states.
In particular, HIS must be represented with three titration states:  two neutral tautomers and one positively charged state.
\revision{To accommodate this increase in states, we represented the neutral states of a HIS residue as two separate residues: HIS${}_\epsilon$ and HIS${}_\delta$.}
If, in the result of the graph-cut algorithm, only one of these states is protonated, then that is the appropriate minimum-energy state.
If both HIS${}_\epsilon$ and HIS${}_\delta$ are protonated, then HIS is fully protonated (in the charged state).
The doubly deprotonated state of HIS is excluded; this exclusion is enforced by an infinite interaction energy between the HIS${}_\epsilon$ and HIS${}_\delta$ states.
{The energetic differences between HIS${}_\epsilon$ and HIS${}_\delta$ are calculated in PDB2PKA and currently only include electrostatic interactions with the environment -- although it is known that these tautomers are not energetically equivalent in isolation \cite{Bhattacharya1997}.}

In order to separate each HIS residue into two separate residues we must define the interaction energy between a separated HIS residue and a non-HIS residue, based on the energies of the non-separated case. We also must define the interaction between two separated HIS residues.
The simplest case involves interactions of HIS with a non-HIS residue.
\revision{The details for constructing the associated energy graph are given in Supporting Information.}

\subsection{Calculation details}

\revision{The $\theta_p$ values were calculated using Eqs.~\ref{eqn:theta2} or \ref{eqn:theta3} based on the lowest-energy state obtained from the graph-cut method.}
The graph-cut algorithm was written in {Python} and executed on desktop computer running Macintosh OSX 10.6.8, with two 2.8 GHz 4 Core Xeon E5462 x2 processors with 20GB RAM.
The resulting \pkatext{} values were compared against the Monte Carlo method in PDB2PKA.
Both algorithms were run to calculate titration curves using pH values from 0 to 20 in increments of 0.1.

Test proteins were selected from the PDB by identifying all entries with a single chain, no ligands or modified amino acids, and resolutions below 2.0 \AA.
The resulting list was then refined by processing through WHATCHECK \cite{WHATCHECK}; proteins with errors or excessive warnings were removed from the list.
Finally, these proteins were run through the PDB2PKA software.
Proteins with excessive energies (typically due to unresolved steric clashes or non-converged electrostatics calculations) were removed from the list.
The final list of 82 proteins is provided in the Supporting Information iPython notebook and shows the distribution of proteins with between 11 and 45 titratable residues.
We calculated titration curves for each of the titratable residues in 87 different proteins, resulting in 2337 predictions for each algorithm.

Recall that in the case of non-submodular interaction energies, which is typically our setting, we may not be able to label all residues using the graph-cut algorithm.
In the cases where we had a significant number of unlabeled residues following the graph-cut method we needed to perform additional optimization.
\revision{If there are fewer than (or equal to) 20 unassigned residues, we do a brute force search to find the minimum energy state.
The ``brute force'' approach enumerates the $2^n$ (for $n$ unlabeled residues) possibilities to identify the one with the lowest energy.
If there are more than 20, we perform a Monte Carlo optimization sampling a subset of the $2^n$ states.}

\section{Results} \label{sec:results}

\begin{figure}
	\begin{center}
		\includegraphics[keepaspectratio=true,width=0.95\linewidth]{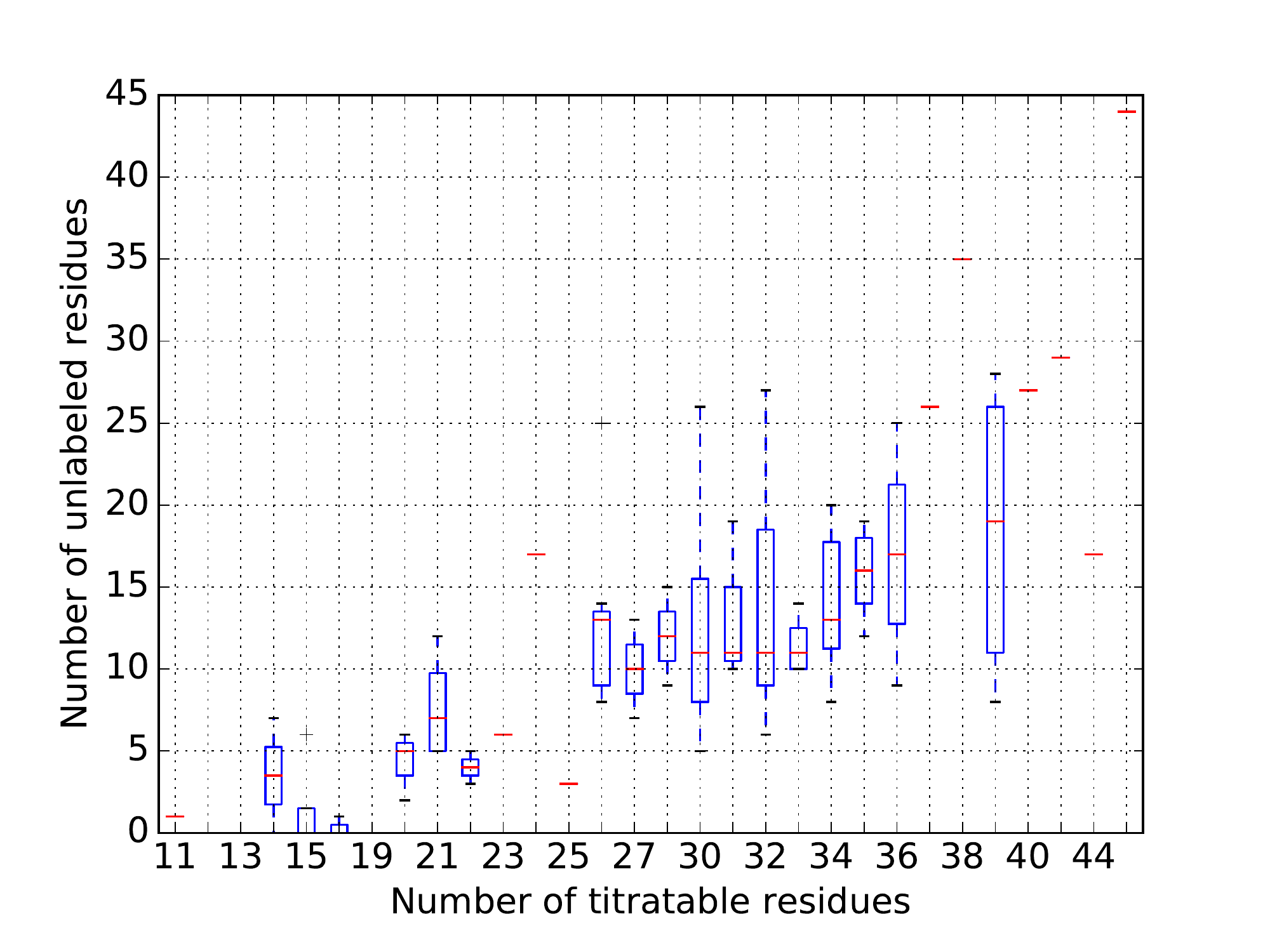}
	\end{center}
	\caption{Distribution (over pH values and proteins) of residues left unlabeled, and thus needing brute force calculation, after applying the graph-cut algorithm.}
	\label{fig:partOptResults}
\end{figure}

As discussed above, there is a possibility that some residues cannot be assigned titration states due to non-submodular energies.
Figure \ref{fig:partOptResults} shows the number of residues which were not assigned using the graph-cut method as a function of protein size.

\begin{figure}
	\begin{center}
		(A)~\includegraphics[width=0.6\linewidth,keepaspectratio]{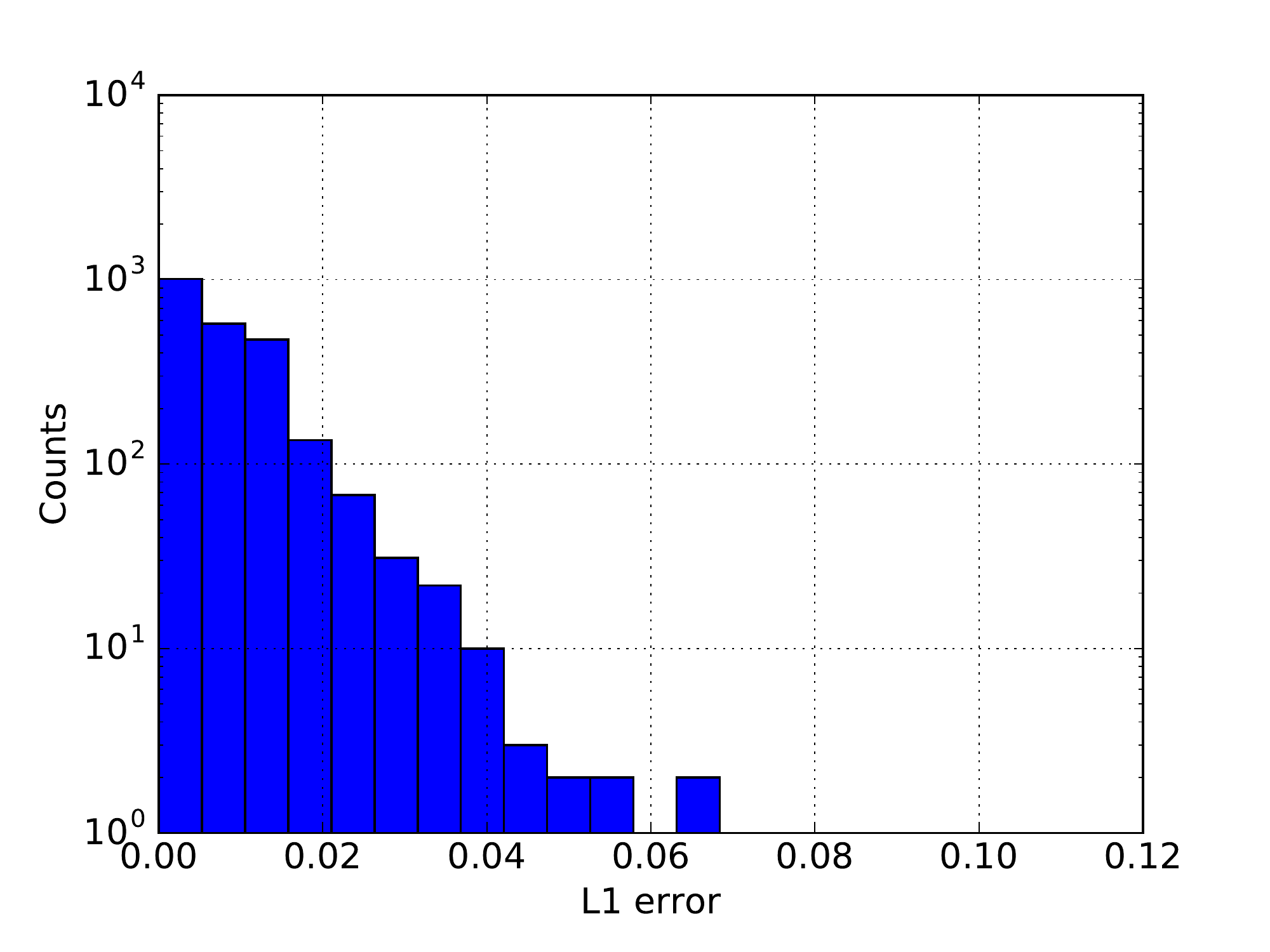}
		(B)~\includegraphics[width=0.6\linewidth,keepaspectratio]{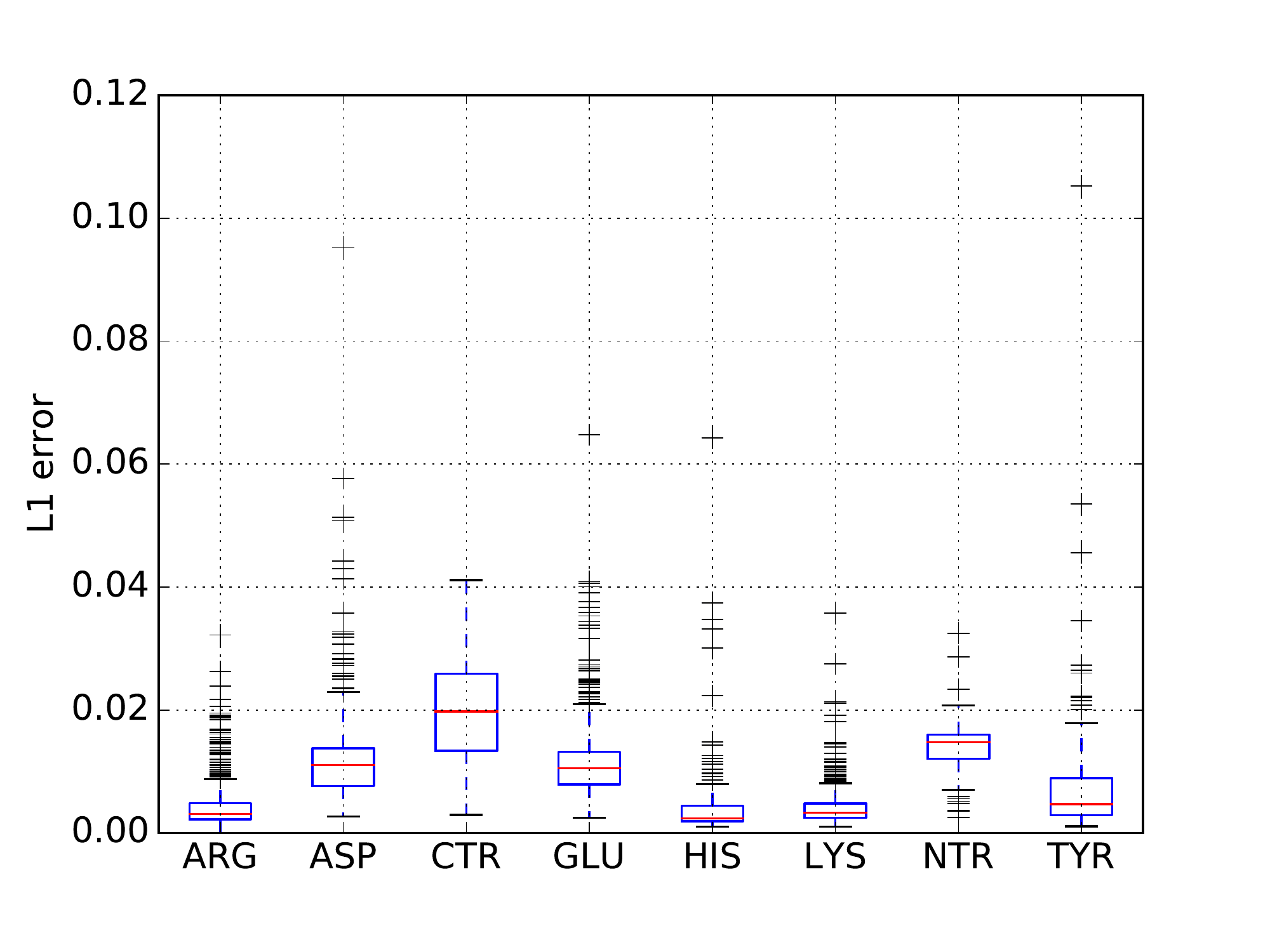}
	\end{center}
	\caption{Comparison of titration curves with differences measured by $\ell_1$ difference between PDB2PKA and graph-cut, as defined in the text (Eq.~\ref{eqn:MAE}).
	(A) Distribution of differences across the 2337 titration curves.
	(B) Distribution of differences for the titratable groups studied in this paper.}
	\label{fig:titrCompare}
\end{figure}

The graph cut and PDB2PKA titration curves were compared by the mean absolute difference of the titration probability integrated over the pH range:
\begin{equation}
	\left\| e \right\|_{\ell_1} = \frac{1}{20} \int_0^{20} \left| p_{\text{PDB2PKA}} - p_{\text{Graph cut}} \right| d \text{pH}
	\label{eqn:MAE}
\end{equation}
with the results shown in Fig.~\ref{fig:titrCompare}.
Results for the mean-squared differences are also provided in Supporting Information.
Most titration curves show high levels of agreement with less than 5\% error.
The three curves with the worst agreement (4PGR TYR 167: 11\% error, 4PGR ASP 195: 10\% error, and 3IDW GLU 51: 7\% error) are shown in the iPython notebook in Supporting Information.
\revision{Further analysis of the few residues which showed large deviation between the PDB2PKA and graph cut titration curves showed very large variations in interaction energies.
In particular, the set of all interaction energies between the residue in question and all other residues show a much wider range of energy values than other better-behaved residues.
We believe that the significant energy outliers confound probabilistic Monte Carlo sampling and optimization and lead to the poor agreement for these few cases.}

\begin{figure}
	\begin{center}
		\includegraphics[keepaspectratio=true,width=0.95\linewidth]{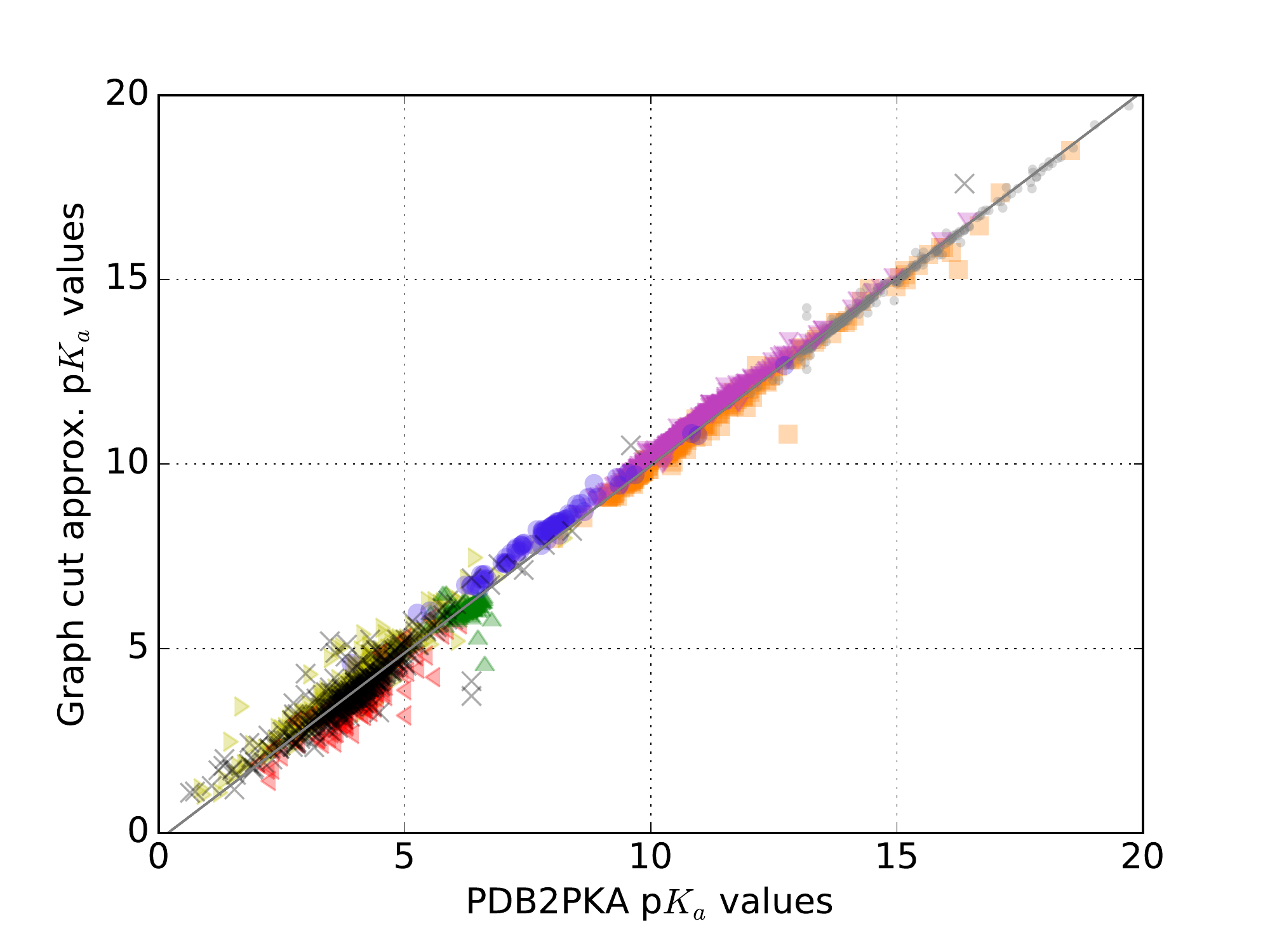}
	\end{center}
	\caption{Comparison of \pkatext{} values calculated with the graph-cut method and with PDB2PKA.
	$\circ$: arginine, $\square$: aspartate, $+$: C-terminus, $\times$: glutamate, $*$: HIS${}_\epsilon$, $\diamond$: HIS${}_\delta$, $\triangle$: lysine, $\triangledown$: N-terminus, $\triangleleft$: tyrosine.
	Line shows linear fit with $p < 0.0001$.}
	\label{fig:pkaCompare}
\end{figure}

The titration curves were used to derive \pkatext{}s by locating the pH values where the curves crossed 0.5.
This approach was used in lieu of Henderson-Hasselbalch fitting because of the coupled titration events observed in several proteins.
Our simpler approach is primarily intended to compare the computational methods to each other -- rather than generating \pkatext{} values for comparison with experiment.
Figure \ref{fig:pkaCompare} compares the \pkatext{} values calculated by the graph-cut and PDB2PKA methods.
The \pkatext{} values are strongly correlated with a Pearson $r^2 = 0.996$, a slope of 1.01, and an intercept of $-0.19$.
\revision{A comparison of \pkatext{} \emph{shifts}\footnote{A \pkatext{} shift is the difference between the observed \pkatext{} value and a model \pkatext{} for that residue type.} is presented in Supporting Information.}

\section{Discussion} \label{sec:discussion}

Our graph-based approach solves an optimization problem to determine the titration state with the lowest energy at a particular pH.
This a potential source of error in our prediction of \pkatext{} values and titration curves because the \revision{titration probabilities} should be calculated as thermodynamic averages over the entire ensemble of states.
However, Figures \ref{fig:titrCompare} and \ref{fig:pkaCompare} show that the results from our optimization approach and ensemble averages (from PDB2PKA) are very similar.
This suggests a scarcity of low-energy states surrounding the titration transitions and is directly related \revision{to} the accessibility of alternate charge states by thermal fluctuations.
This phenomenon was described by Kirkwood and Shumaker \cite{KiSh52} and continues to be an active area of theoretical and computational investigation \cite{PhysRevE.91.022715}.
The effects of thermal fluctuations are expected to be largest when the pH is close to the \pkatext{} of a site \cite{KiSh52, Swails2012, Itoh2011, Stern2007}.
The close agreement between the optimization and Monte Carlo results in Figures \ref{fig:titrCompare} and \ref{fig:pkaCompare} indicate that such fluctuations do not play a major role in the system we studied.
However, future work will explore the possibility of using the graph-cut optimization as input to Monte Carlo sampling around the energy minimum to efficiently sample only the most relevant fluctuations.

As described above, residues that violate the modularity condition for titration site interactions  are not labeled by the graph cut algorithm and must be optimized by either brute force or Monte Carlo methods.
The iPython notebook provided in Supporting Information illustrates the impact of unlabeled residues on execution time.
Future research will focus on better understanding the influence of protein structure and energetics on this submodularity and to use more sophisticated optimization methods on the residues not labeled in the graph cut optimization procedure.

\section{Conclusions}

Most current titration state prediction algorithms suffer from either performance or sampling issues when searching over the $\order{2^N}$ titration states associated with $N$ titratable residues in a protein system.
Sampling issues have historically been a major problem for the PDB2PQR/PDB2PKA Monte Carlo approach for sampling titration states as well as other software packages.
This paper presented a new polynomial-time algorithm for optimization of discrete titration states in protein systems.
This algorithm was adapted from image processing and uses techniques from discrete mathematics and graph theory to restate the optimization problem in terms of ``maximum flow-minimum cut'' graph analysis.
The interaction energy graph, a graph in which vertices (amino acids) and edges (interactions) are weighted with their respective energies, is transformed into a flow network in which the value of the minimum cut in the network equals the minimum free energy of the protein, and the cut itself encodes the state that achieves the minimum free energy.
Because of its deterministic nature and polynomial-time performance, this algorithm has the potential to allow for the ionization state of larger proteins to be discovered.

There are several other problems in macromolecular modeling that require optimization over an exponentially large space of states, including inverse protein folding and design \cite{yue_inverse_1992, GuAMaJStL2005} and rotamer sampling/selection \cite{JaTCeDMcJ2006}.
In the future, we plan to extend this work to some of these other applications.
However, the systematic extension of this approach to multi-state systems will be challenging.
For example, in order to adapt this work to rotamer selection we need many more than two labels per amino acid and a more generalizable approach for decomposing pairwise interactions between rotamer states.
In particular, we need a way of creating a flow network that will allow us to handle any number of labels per amino acid.
There has been work in multi-label algorithms for computer vision \cite{BoYVeOZaR2001, CaPHaR2009, VeO1999, VeO2009, PrS2012}, however they all require that the energy functions satisfy submodularity as introduced earlier in this manuscript.
However, algorithms such as $\alpha-$expansion and $\alpha-\beta$-swap which will calculate minimum energy approximations without the need for such restrictions \cite{VeO1999}.
In future work, we plan to combine these approximations with the non-submodular case discussed in the rest of this paper in order to remove dependency on these
energy function conditions.

\section{Supporting Information Available}
In a separate supporting information document we provide more details on graph theory formalism, network flows, the creation of the energy graph, normal form, and energy flow network, and reduction of HIS from one amino acid with three states to two interacting amino acids with two states.
Additionally we provide all of our data and an iPython notebook in order to reproduce our results, and dig further into the analysis of our algorithms.
Finally, we also provide the graph cut Python code which can compute minimum energy states and titration curves using the algorithms described in this paper.

\section*{Acknowledgements}
We gratefully acknowledge NIH grant GM069702 for support of this research, Dr.~Jens Nielsen for his work on PDB2PKA, and the National Biomedical Computation Resource (NIH grant RR008605) for computational support.

\bibliography{pka}

\end{document}